# Spectral adjustment of a powerful random fiber laser

Jun Ye[1], Jiangming Xu[1,2], Hanshuo Wu[1], Jiaxin Song[1], Hanwei Zhang[1,2], Jian Wu[1,2], Pu Zhou[1,2,*]

**Abstract:** Random fiber laser (RFL) based on random distributed feedback and Raman gain has earned much attention in recent years. In this presentation, we demonstrate a powerful linearly polarized RFL with spectral adjustment, in which the central wavelength and the linewidth of the spectrum can be tuned independently through a bandwidth-adjustable tunable optical filter (BA-TOF). As a result, the central wavelength can be continuously tuned from 1095 to 1115 nm, while the full width at half-maximum (FWHM) linewidth has a maximal tuning range from ~0.6 to more than 2 nm. This laser provides a flexible tool for scenes where the temporal coherence property accounts, such as coherent sensing/communication and nonlinear frequency conversion. To the best of our knowledge, this is the first demonstration of a high power linearly polarized RFL with both wavelength and linewidth tunability.

## 1. Introduction

In 2010, Turitsyn et al. [1] demonstrated a new type of random fiber laser (RFL), which makes use of Rayleigh scattering (RS) that produces random distributed feedback (RDFB) and stimulated Raman scattering (SRS) that provides amplification, the most obvious features of RFL are cavity-free and mode-less [2]. Over recent years, RFL has drawn a great deal of attention, gradually leads to the realization of high power [3, 4], narrow-linewidth [5-7], multi-wavelength [8, 9] and linearly polarized operation [7, 10, 11]. Thanks to the good laser performance and relative simplicity of implementation, RFLs have attracted a large variety of applications, such as frequency doubling to the visible [12], sensing and telecommunication [13-15], pump source in mid-infrared laser and supercontinuum light source [16-19], and stable seed for high power fiber master oscillator power amplifier (MOPA) [20, 21].

For better understanding the physical mechanism of random lasing and widening the application range in optical communication, sensing, secure transmission and other fields, various kinds of RFL schemes with wavelength tunability have been demonstrated [22-28]. However, the linewidth parameter of laser spectrum, which is related to the inherent temporal property of a laser, has not yet been reported. In this presentation, we demonstrate a powerful RFL with both wavelength and linewidth tunability for the first time to our best knowledge. In the proof-of-principle experiment, the central wavelength can be continuously tuned from 1095 to 1115 nm, while the maximal tuning range of the full width at half-maximum (FWHM) linewidth is from ~0.6 to more than 2 nm. Moreover, the output power of 1102.5-1112.5 nm reaches ~23 W with polarization extinction ratio (PER) value >20 dB.

## 2. Experimental setup

The schematic diagram of the experimental setup is shown in figure 1. The pumping is provided by a linearly polarized all-fiber MOPA source centered at 1055 nm, which can deliver an output power up to 46.9 W with the FWHM linewidth <1.2 nm. A Polarization maintaining isolator (PM ISO) is followed to protect the pump source. The linearly polarized pump radiation is injected into a piece of 450-m-long PM passive fiber via the 1060 nm port of a high power PM wavelength division multiplexer (WDM), the 1110 nm port of the PM WDM is spliced to a PM coupler with coupling ratio of 50/50, which forms a fiber loop mirror (FLM) by splicing its output ports. To obtain wavelength-tunable and linewidth-adjustable random emission, a BA-TOF with 1-m-long pigtailed fiber is placed into the FLM. In order to monitor the operation of the BA-TOF, a PM tapper with coupling ratio of 1/999 is also placed into the FLM. In addition, all the end facets are cleaved at an angle of 8° to eliminate unexpected Fresnel reflection.

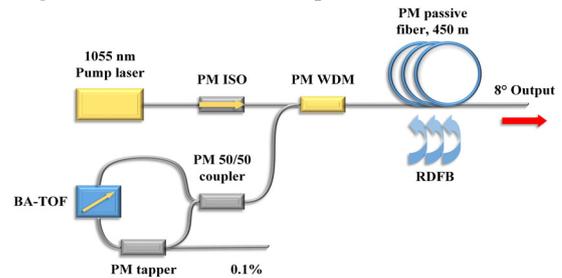

Fig. 1. Schematic diagram of the experimental setup. PM ISO: Polarization maintaining isolator, PM WDM: Polarization maintaining wavelength division multiplexer, BA-TOF: Bandwidth-adjustable tunable optical filter, RDFB: Random distributed feedback.

## 3. Results and discussion

### A. Wavelength Tunability

Inserting the BA-TOF and fixing its passband at the minimum value, we first explored the wavelength tunability of the RFL. As shown in figure 2(a), the normalized output spectra have a tuning range of ~20 nm (from 1095 to 1115 nm).

1 College of Optoelectronic Science and Engineering, National University of Defense Technology, Changsha 410073, China

2 Hunan Provincial Collaborative Innovation Center of High Power Fiber Laser, Changsha 410073, China.

* Corresponding author: zhoupu203@163.com

Figure 2(b) shows the output power of the first order Stokes dependence on the operating wavelength with the pump power of 34.1 W, 40.6 W and 46.9 W. An apparent similarity between the power variation curve and the Raman gain profile can be observed, where the output powers reach the highest value at ~1107.5 nm and ~1112.5 nm, which correspond to the two peaks of the Raman gain with frequency shifts of ~13.2 THz and ~14.6 THz. Furthermore, the maximal output powers of 1102.5-1112.5 nm reach more than 20 W with power fluctuation less than 1 dB. As a result, due to the available bandwidth of the Raman gain, the wavelength tuning range is confined to ~20 nm, which is a typical value in tunable RFLs with only Raman gain [22-24].

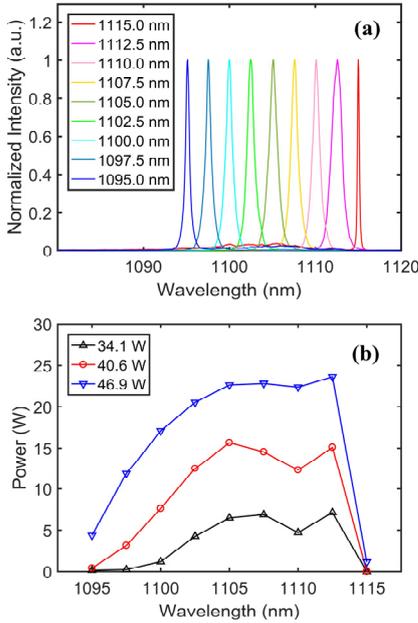

Fig. 2. (a) Normalized tunable output spectra. (b) Output power of the first order Stokes as a function of wavelength with the pump power of 34.1 W, 40.6 W and 46.9 W.

### B. Linewidth Tunability

On the other hand, fixing the central wavelength, we analyzed the linewidth tunability of the RFL. Take the case of operating at 1107 nm, the linewidth-adjustable spectra of the output beam are depicted in figure 3. The output spectrum shows a clear peak with relatively long tails. Via increasing the passband of the BA-TOF, the output spectra can gradually broaden, where the FWHM linewidths can be continuously tuned from ~0.61 nm to ~2.03 nm (the corresponding 10 dB linewidth varies from ~1.97 nm to ~4.02 nm). An interesting phenomenon can be observed when further increasing the passband of the BA-TOF, that is the abrupt change of the FWHM linewidth. When adjusting the passband to 5.4 nm, the FWHM linewidth narrows to ~1.48 nm. However, when the passband is increased to 6.2 nm, the FWHM linewidth immediately broadens to ~5.24 nm.

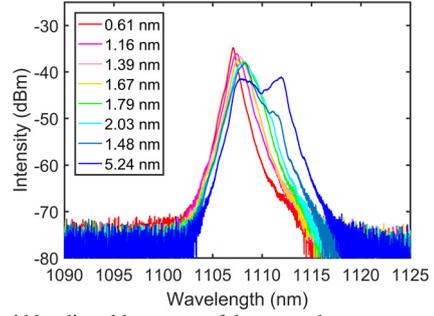

Fig. 3. Linewidth-adjustable spectra of the output beam.

The reason could be attributed to the working principle of the BA-TOF, whose passband is increased via broadening the aperture inside to let the longer wavelength through. Thus, due to the high Raman gain at frequency shift of ~14.6 THz, the subcomponent near 1112 nm will start to radiate immediately as long as this wavelength is covered by the passband of the BA-TOF. As a result, the subcomponent of ~1107 nm and the subcomponent of ~1112 nm are simultaneously present in the output spectrum, thus leading to the abrupt change of the FWHM linewidth. In fact, the concept of FWHM linewidth is not universally applicable for single-peak spectrum and multi-peak spectrum. Figure 4 shows the output power of the random emission while tuning the linewidth (measured at 1107 nm with 46.9 W pump). The laser output power reaches ~23 W and keeps almost constant with a variation of ~8% in the tuning range 0.6-2.0 nm.

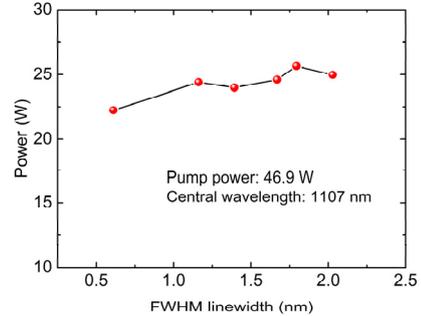

Fig. 4. Output power as a function of the FWHM linewidth.

Finally, the FER value of the linearly polarized RFL is analyzed briefly. As a result, with the pump power grows and well above the threshold, we found that the PER value keeps nearly the same, slightly varying between 21 and 22 dB.

### 4. CONCLUSION

In summary, we reported a powerful linearly polarized RFL with spectral adjustment, especially the linewidth tunability, which is related to the inherent temporal property of a RFL, is explored for the first time, as far as we know. As a result, the central wavelength can be continuously tuned from 1095 to 1115 nm, while the FWHM linewidth has a maximal tuning range from ~0.6 to more than 2 nm. Moreover, the output power of 1102.5-1112.5 nm reaches ~23 W with PER value >20 dB.